\documentclass{article}

\PassOptionsToPackage{numbers, compress}{natbib}
\usepackage[preprint]{neurips_2023}

\usepackage{threeparttable}
\usepackage[utf8]{inputenc} 
\usepackage[T1]{fontenc}    
\usepackage{hyperref}       
\usepackage{url}            
\usepackage{booktabs}       
\usepackage{amsfonts}       
\usepackage{nicefrac}       
\usepackage{microtype}      
\usepackage{xcolor}         
\usepackage{amsmath}
\usepackage{amssymb}
\usepackage{mathtools}
\usepackage{amsthm}
\usepackage{multirow}
\usepackage{wrapfig}
\usepackage[capitalize,noabbrev]{cleveref}
\usepackage{graphicx}
\usepackage{subfigure} 
\usepackage{float}
\usepackage{algorithm} 
\usepackage{algorithmic}
\usepackage{textcomp} 
\usepackage{listings} 
\usepackage{enumitem}
\usepackage{utfsym}
\usepackage{times}
\usepackage{latexsym}
\usepackage{fancyvrb}
\usepackage{CJKutf8}
\usepackage{tabularx}

\newcommand{\commentout}[1]{}
\newcommand{\chn}[1]{\begin{CJK*}{UTF8}{gbsn}{#1}\end{CJK*}}

\title{FireRedASR2S: A State-of-the-Art Industrial-Grade All-in-One Automatic Speech Recognition System}

\author{%
\centering
  \begin{tabular}{c}
Kaituo Xu, Yan Jia, Kai Huang, Junjie Chen, Wenpeng Li, Kun Liu\\
Feng-Long Xie, Xu Tang, Yao Hu\\
{\normalfont Super Intelligence Team, Xiaohongshu Inc.}
\end{tabular}%
}

\begin{document}
\maketitle
\begin{abstract}
We present FireRedASR2S, a state-of-the-art industrial-grade all-in-one automatic speech recognition (ASR) system. It integrates four modules in a unified pipeline: ASR, Voice Activity Detection (VAD), Spoken Language Identification (LID), and Punctuation Prediction (Punc). All modules achieve SOTA performance on the evaluated benchmarks:

\textbf{FireRedASR2}: An ASR module with two variants, FireRedASR2-LLM (8B+ parameters) and FireRedASR2-AED (1B+ parameters), supporting speech and singing transcription for Mandarin, Chinese dialects and accents, English, and code-switching. Compared to FireRedASR, FireRedASR2 delivers improved recognition accuracy and broader dialect and accent coverage. FireRedASR2-LLM achieves 2.89\% average CER on 4 public Mandarin benchmarks and 11.55\% on 19 public Chinese dialects and accents benchmarks, outperforming competitive baselines including Doubao-ASR, Qwen3-ASR, and Fun-ASR.

\textbf{FireRedVAD}: An ultra-lightweight module (0.6M parameters) based on the Deep Feedforward Sequential Memory Network (DFSMN), supporting streaming VAD, non-streaming VAD, and multi-label VAD (mVAD). On the FLEURS-VAD-102 benchmark, it achieves 97.57\% frame-level F1 and 99.60\% AUC-ROC, outperforming Silero-VAD, TEN-VAD, FunASR-VAD, and WebRTC-VAD.

\textbf{FireRedLID}: An Encoder-Decoder LID module supporting 100+ languages and 20+ Chinese dialects and accents. On FLEURS (82 languages), it achieves 97.18\% utterance-level accuracy, outperforming Whisper and SpeechBrain.

\textbf{FireRedPunc}: A BERT-style punctuation prediction module for Chinese and English. On multi-domain benchmarks, it achieves 78.90\% average F1, outperforming FunASR-Punc (62.77\%).

To advance research in speech processing, we release model weights and code at \url{https://github.com/FireRedTeam/FireRedASR2S}.
\end{abstract}

\section{Introduction}
\label{sec:intro}

Automatic speech recognition (ASR) has advanced rapidly with end-to-end modeling, large-scale training, and the integration of large language models (LLMs) \cite{xu2025fireredasr,shi2026qwen3,an2025fun,seedasr2024,tian2026dllm,mu2025efficient,an2024funaudiollm,chu2024qwen2,chu2023qwen,wu2023decoder,rubenstein2023audiopalm,li2023prompting,wang2023slm,pan2023cosmic,yu2024connecting,chen2024salm,lakomkin2024end,geng2024unveiling,ma2024embarrassingly}.
However, practical deployment in real-world applications typically requires more than a standalone ASR model.
Real-world audio often contains long-form recordings, silence and non-speech regions, background music, singing, multilingual speech, and Chinese dialects and accents (hereafter referred to as dialects for brevity).
To deliver reliable transcription in such conditions, a complete pipeline is needed, including voice activity detection (VAD) for segmentation, spoken language identification (LID) for routing and tagging, and punctuation prediction (Punc) for readable outputs.

In practice, many systems are built by assembling modules from heterogeneous sources (e.g., separate VAD/LID/ASR/Punc toolkits or cloud services).
Such pipelines frequently suffer from inconsistent interfaces, limited reproducibility, and complex error propagation.
Moreover, some components rely on weak or indirect supervision (e.g., VAD trained from ASR forced alignment), which may degrade robustness under challenging acoustic conditions.
These limitations motivate an open-source, industrial-grade all-in-one ASR system with strong performance, clear modularization, and comprehensive evaluation.

In this technical report, we present \textbf{FireRedASR2S}, a state-of-the-art (SOTA), all-in-one ASR system integrating four modules: \textbf{FireRedASR2} for ASR, \textbf{FireRedVAD} for VAD and multi-label VAD (mVAD), \textbf{FireRedLID} for multilingual and dialect LID, and \textbf{FireRedPunc} for punctuation prediction.
The suffix \textbf{2S} denotes the \textbf{2}nd-generation FireRedASR, expanded into an all-in-one ASR \textbf{S}ystem.
FireRedASR2S provides a unified pipeline from waveform to structured transcription outputs, while allowing each module to be deployed independently.

FireRedASR2 builds upon our previous FireRedASR \cite{xu2025fireredasr} models with minimal architectural changes.
Compared to FireRedASR, FireRedASR2 improves recognition accuracy and expands coverage to a broader range of Chinese dialects, primarily by scaling supervised training data to approximately 200k hours with broader domain, language, and dialect diversity.
FireRedVAD is trained on thousands of hours of high-quality human-annotated acoustic event data, providing reliable segmentation under diverse acoustic conditions.
FireRedLID is implemented as an Encoder-Decoder-based \cite{bahdanau2016end,chan2016listen} model initialized from the FireRedASR2-AED encoder and performs hierarchical language and dialect prediction.
FireRedPunc adopts a BERT-style encoder \cite{devlin2019bert} initialized from LERT \cite{cui2022lert} and is trained on large-scale multi-domain Chinese and English corpora.

Our main contributions are:
\begin{itemize}[leftmargin=1.2em]
\item \textbf{All-in-one open-source system}: We release an integrated ASR pipeline with unified interfaces and modular deployment.
\item \textbf{Improved ASR accuracy and broader dialect coverage}: Building upon FireRedASR, FireRedASR2 improves recognition accuracy and expands support for Chinese dialects, achieving strong results on 24 public test sets.
\item \textbf{Robust segmentation from human-labeled events}: FireRedVAD provides strong multilingual VAD performance and is trained using high-quality human-annotated event data rather than forced-alignment-derived supervision.
\item \textbf{Hierarchical multilingual and dialect LID}: FireRedLID supports 100+ languages and 20+ Chinese dialects with a compact two-token decoding formulation.
\item \textbf{Effective punctuation prediction}: FireRedPunc achieves strong results on multi-domain Chinese and English punctuation benchmarks.
\end{itemize}

The remainder of this report is organized as follows.
\Cref{sec:system_overview} presents the system overview.
\Cref{sec:asr2,sec:vad,sec:lid,sec:punc} describe each module.
\Cref{sec:eval} reports evaluation results.
\Cref{sec:discussion} discusses key design choices and limitations, and \Cref{sec:conclusion} concludes the report.

\section{FireRedASR2S: System Overview}
\label{sec:system_overview}
FireRedASR2S is an industrial-grade, all-in-one ASR system that integrates four modules---FireRedVAD (\Cref{sec:vad}), FireRedLID (\Cref{sec:lid}), FireRedASR2 (\Cref{sec:asr2}), and FireRedPunc (\Cref{sec:punc})---into a unified pipeline.
The system is designed in a modular manner: each component can be used independently, while the default configuration forms an end-to-end transcription pipeline that handles diverse acoustic conditions (speech, singing, music, and non-speech) as well as multilingual and Chinese dialect scenarios, and produces structured outputs including punctuated text, timestamps, confidence scores, and language labels.

\begin{figure}[!ht]
\centering
\includegraphics[width=\linewidth]{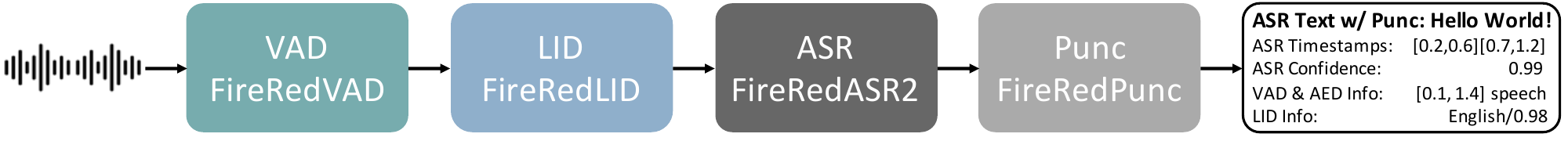}
\caption{Overview of FireRedASR2S. The input waveform is processed sequentially by FireRedVAD (VAD), FireRedLID (LID), FireRedASR2 (ASR), and FireRedPunc (Punc) to produce structured transcription outputs, including punctuated text, timestamps, confidence scores, and language labels.}
\label{fig:system_overview}
\end{figure}

\noindent\textbf{Pipeline}:
As illustrated in \Cref{fig:system_overview}, FireRedASR2S processes an input waveform through four stages.
First, FireRedVAD detects voice segments on the original audio timeline and filters out non-voice regions to improve robustness on long-form audio.
Second, FireRedLID predicts an utterance-level language label for each detected segment and, when applicable, a Chinese dialect label.
Third, FireRedASR2 transcribes each segment into text and returns an ASR confidence score; when using FireRedASR2-AED, it additionally provides token- and word-level timestamps.
Finally, FireRedPunc restores punctuation for the ASR output to improve readability and downstream usability.

\noindent\textbf{Structured outputs}:
FireRedASR2S returns structured outputs containing (1) the final transcription text, (2) a list of sentence-level segments with start/end timestamps, recognized text, ASR confidence, and optional language labels with confidence, and (3) VAD segmentation results.
When ASR timestamping is enabled, the system can further derive sentence-level timestamps by leveraging punctuation prediction.
All timestamps are reported on the original waveform timeline.

\noindent\textbf{Modularity}:
Although FireRedASR2S is designed as an end-to-end pipeline, each module can be deployed as a standalone component (e.g., VAD-only segmentation, LID-only routing, ASR on pre-segmented audio, or punctuation on plain text).
This modular design enables flexible deployment and independent iteration of each component.

\section{FireRedASR2: Automatic Speech Recognition}
\label{sec:asr2}

We summarize the key components of FireRedASR2 here and highlight the incremental updates. For detailed specifications of the Conformer and Transformer blocks, the Encoder-Adapter-LLM training procedure, and the optimization strategies, we refer readers to the FireRedASR technical report \cite{xu2025fireredasr}.

FireRedASR2 comprises two variants: FireRedASR2-AED and FireRedASR2-LLM. FireRedASR2-AED follows the conventional Attention-based Encoder-Decoder architecture \cite{bahdanau2016end,chan2016listen}, whereas FireRedASR2-LLM is built on the Encoder-Adapter-LLM architecture \cite{xu2025fireredasr,shi2026qwen3,an2025fun,seedasr2024,wu2023decoder,geng2024unveiling,ma2024embarrassingly} that leverages the power of LLM for ASR. Both models share similar input feature processing and acoustic encoding strategies but differ in their approaches to token sequence modeling. FireRedASR2-AED additionally supports token-level timestamps and utterance-level confidence scores. Word-level timestamps are obtained by post-processing token timestamps (e.g., merging English subword units into words).

\begin{figure*}[!ht]
\centering
\includegraphics[width=\linewidth]{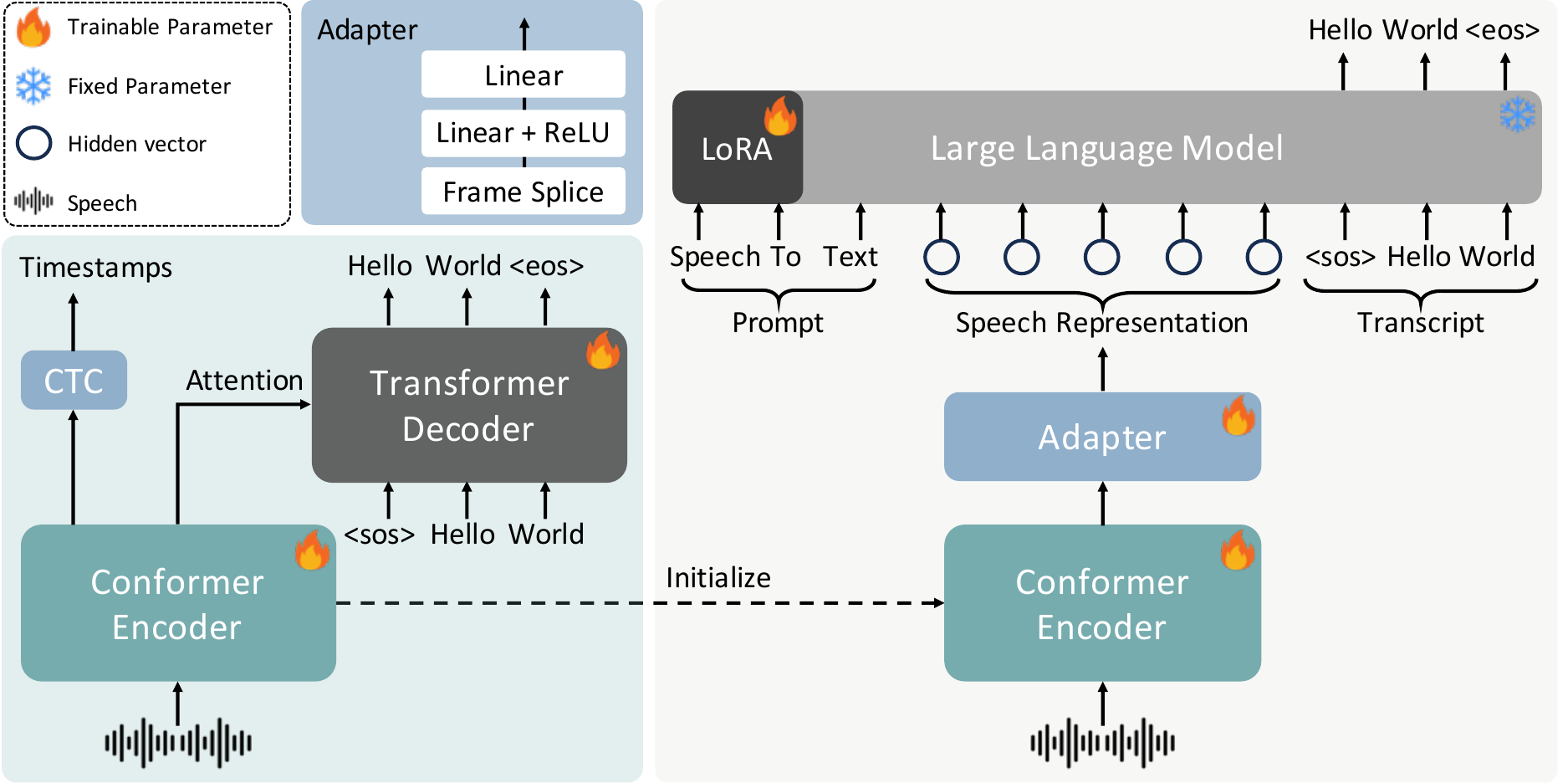}
\caption{Architecture of FireRedASR2-AED (bottom left), FireRedASR2-LLM (right), and Adapter.}
\label{img:llm_asr}
\end{figure*}

\subsection{FireRedASR2-AED: Attention-based Encoder-Decoder ASR model}
\label{sec:asr2_aed}
The overall architecture of FireRedASR2-AED is illustrated in Figure \ref{img:llm_asr} (bottom left).

\noindent\textbf{Architecture}: FireRedASR2-AED adopts an end-to-end ASR architecture that follows the Conformer-based Encoder and Transformer-based Decoder design of FireRedASR-AED  \cite{xu2025fireredasr,gulati2020conformer,vaswani2017attention}. The Encoder begins with convolutional subsampling to reduce frame rate and is followed by stacked Conformer blocks. The Decoder is a standard Transformer-based Decoder attending to the Encoder states to generate token sequences with a cross-entropy objective.
Unless otherwise specified, architectural hyperparameters and training recipes follow FireRedASR-AED (model size L) \cite{xu2025fireredasr}.

\noindent\textbf{Training Data}: Compared to FireRedASR, the primary update of FireRedASR2 is the expansion of supervised training data from 70k hours to approximately 200k hours. The corpus covers Mandarin, English, Chinese dialects, code-switching, speech and singing, as well as non-speech audio. We attribute the performance gains and improved generalization of FireRedASR2 primarily to this larger and more diverse training corpus.

\noindent\textbf{Input Features}: We use 80-dimensional log Mel filterbank (FBank) features extracted from 25ms windows and 10ms frame shifts, followed by global mean and variance normalization (CMVN). 

\noindent\textbf{Tokenization}: FireRedASR2-AED adopts a mixed tokenization strategy: Chinese characters for Chinese text and token-level byte-pair encoding (BPE) \cite{sennrich2015neural} tokens for English text. Compared to FireRedASR-AED, FireRedASR2-AED uses an updated vocabulary size of 8,667 to better cover multilingual and dialect scenarios.

\subsubsection{Confidence estimation from decoder probabilities}
\label{sec:asr2_conf}
FireRedASR2-AED returns an utterance-level confidence score to indicate the reliability of the transcription. This score is derived from the decoder's token probabilities. Specifically, we extract the per-token posterior probabilities (i.e., softmax outputs) along the 1-best hypothesis produced by beam search, excluding special tokens. These token-level probabilities are then aggregated into a single sequence-level score, typically formulated as the geometric mean of the valid tokens. To improve reliability in practical deployments, this raw aggregated score can be further refined using heuristic strategies (e.g., filtering out statistical outliers or applying confidence clipping). Finally, this sequence-level score can be used for downstream filtering, ranking, or UI display.

\subsubsection{Post-hoc CTC branch for timestamps}
\label{sec:asr2_ctc}
A key update in FireRedASR2-AED is the support of timestamps via an additional CTC \cite{graves2006connectionist} branch attached to the encoder. After the base AED model (Conformer encoder + Transformer decoder) is fully trained, we add a lightweight CTC projection head on top of the encoder outputs and train it post-hoc by freezing the encoder and decoder and optimizing only the CTC branch with the standard CTC objective. 
The CTC head is implemented as a linear projection from encoder hidden states to logits, and the CTC vocabulary is identical to the AED vocabulary to enable forced alignment between CTC posteriors and AED-decoded tokens.
This design preserves the recognition accuracy of the base AED model while enabling alignment-based timestamp prediction.

During inference, we first decode the token sequence using the AED decoder (beam search). We then compute frame-level CTC logits from encoder states and perform CTC forced alignment between the CTC logits and the AED-decoded token sequence (with blank id set to 0). The frame-level alignment is converted into token-level start/end times according to the encoder subsampling rate.

For the final system output, we provide word-level timestamps by post-processing token timestamps. Specifically, we merge timestamps of subword units into words by grouping the corresponding BPE tokens and taking the minimum start time and maximum end time within each merged word. For Chinese, we treat each character token as a word unit.

\subsection{FireRedASR2-LLM: Encoder-Adapter-LLM-based ASR model}
\label{sec:asr2_llm}
FireRedASR2-LLM is also an end-to-end ASR model and follows the Encoder-Adapter-LLM framework of FireRedASR-LLM \cite{xu2025fireredasr}. The model consists of: (1) a Conformer-based audio Encoder that transforms acoustic features into high-level representations, (2) a lightweight Adapter that maps encoder outputs into the embedding space of a pretrained text LLM \cite{qwen2}, and (3) an autoregressive LLM that performs next-token prediction to generate the transcript. The overall architecture of FireRedASR2-LLM is illustrated in Figure \ref{img:llm_asr} (right).

FireRedASR2-LLM employs the same training data, input features and processing methods as FireRedASR2-AED. The encoder of FireRedASR2-LLM is initialized with pre-trained weights from the encoder of FireRedASR2-AED.

The key change from FireRedASR-LLM to FireRedASR2-LLM is the expanded 200k hours supervised training corpus described in Section~\ref{sec:asr2_aed}. The architecture and the training strategy otherwise remain the same as FireRedASR-LLM. We refer readers to \cite{xu2025fireredasr} for details such as prompt formatting, parameter-efficient LLM adaptation, and decoding configurations.

\subsection{Summary of differences from FireRedASR}
\label{sec:asr2_diff}
Table~\ref{tab:asr2_vs_asr1} summarizes the major differences between FireRedASR2 and FireRedASR. Overall, FireRedASR2 retains the proven model designs in FireRedASR, while improving generalization via a larger and more diverse training corpus and enabling timestamp generation via a post-hoc CTC branch in the AED variant.

\begin{table}[!ht]
\caption{Key updates from FireRedASR to FireRedASR2.}
\label{tab:asr2_vs_asr1}
\centering
\begin{tabular}{l|l|l}
\toprule
\textbf{Item} & \textbf{FireRedASR} & \textbf{FireRedASR2} \\
\midrule
Training data & \textasciitilde 70k hours & \textasciitilde 200k hours \\
Vocab size (AED) & 7,832 & 8,667 \\
Timestamps (AED) & Not Supported & Supported \\
\bottomrule
\end{tabular}
\end{table}

\section{FireRedVAD: Voice Activity Detection}
\label{sec:vad}

FireRedVAD provides robust segmentation for downstream ASR in real-world audio, where speech may co-exist with singing, background music, and various non-speech acoustic events.
Unlike many industrial VAD solutions that rely on ASR forced-alignment signals and are trained primarily on ASR corpora, FireRedVAD is trained on high-quality human-annotated acoustic event data, enabling more reliable detection under complex acoustic conditions.

FireRedVAD includes three DFSMN-based models: (1) a \textbf{non-streaming VAD} model for offline segmentation, (2) a \textbf{streaming VAD} model for low-latency online segmentation, and (3) a \textbf{non-streaming multi-label VAD (mVAD)} model for acoustic event recognition.

\subsection{Tasks and label definitions}
\label{sec:vad_task}

\noindent\textbf{Multi-label VAD (mVAD)}:
mVAD is formulated as a frame-level multi-label classification task over three event posteriors:
\textit{speech}, \textit{singing}, and \textit{music}.
The mVAD model outputs an independent posterior probability for each event, and event segments are obtained via event-wise post-processing.

\noindent\textbf{Voice Activity Detection (VAD)}:
VAD is formulated as a frame-level binary classification task to predict \textit{voice} versus \textit{non-voice}.
We define \textit{voice} as the union of \textit{speech} and \textit{singing}, and \textit{non-voice} as \textit{music}, \textit{silence}, and \textit{noise}.
This definition matches typical ASR usage in user-generated-content (UGC) scenarios, where singing segments are often processed similarly to speech.

\subsection{Training data}
\label{sec:vad_data}

\noindent\textbf{Human-annotated event corpus}:
We train FireRedVAD using thousands of hours of human-annotated acoustic event data.
Each utterance is annotated with time boundaries for \textit{speech}, \textit{singing}, and \textit{music}.
Unlike common practice of deriving VAD supervision from ASR forced alignment or weak segmentation heuristics, FireRedVAD uses direct human annotations.

\noindent\textbf{Supervision for mVAD and VAD}:
The mVAD model uses the original three-class labels directly.
The VAD models use binary labels derived from the same annotation space by mapping \textit{speech} and \textit{singing} to the positive class and mapping \textit{music}, \textit{silence}, and \textit{noise} to the negative class.
Although the tasks share a related ontology, mVAD and VAD are trained as separate models with task-specific objectives and post-processing criteria.

\subsection{Model architecture}
\label{sec:vad_arch}

\noindent\textbf{Input features}:
FireRedVAD uses the same acoustic features as FireRedASR2 (\Cref{sec:asr2}).

\noindent\textbf{DFSMN backbone}:
All FireRedVAD models adopt a Deep Feedforward Sequential Memory Network (DFSMN) \cite{zhang2018dfsmn,gao2023funasr}, which is effective and efficient for frame-level acoustic classification.
We implement FSMN \cite{zhang2016compact} memory blocks using depthwise 1-D convolutions with dilation to model temporal context, together with residual connections for stable optimization.

\noindent\textbf{Network configuration}:
We use 8 DFSMN blocks followed by one additional feed-forward layer.
The hidden size is 256 and the projection size is 128.
For temporal context, we use look-back order 20 with stride 1.
For non-streaming VAD and mVAD, we use look-ahead order 20 with stride 1 to utilize future context for improved offline segmentation.
For streaming VAD, we use look-ahead order 0 to ensure causal inference.
We apply dropout with rate 0.05.

\noindent\textbf{Model size}: Thanks to the compact design, all three FireRedVAD models are extremely lightweight, each containing only $\sim$0.6M parameters (approximately 2.2 MB in float32 format). This ultra-lightweight footprint ensures minimal memory and computational overhead, making them highly suitable for massive concurrent processing on cloud servers as well as low-resource edge deployment.

\noindent\textbf{Output layer}:
The final classifier is a linear projection from DFSMN states to logits.
VAD uses a one-dimensional output (voice vs.\ non-voice), while mVAD uses a three-dimensional output (speech, singing, and music).
We apply sigmoid activations to obtain posterior probabilities.

\noindent\textbf{Streaming inference}:
To support online VAD, the streaming model maintains a small per-layer cache that stores a fixed-length history required by the FSMN look-back memory.
During inference, the model updates caches incrementally and outputs frame posteriors without reprocessing past audio, enabling low-latency and bounded-memory streaming.

\subsection{Post-processing and segmentation}
\label{sec:vad_post}
The DFSMN models produce frame-level posterior probabilities, which are converted into time segments via a deterministic post-processing pipeline. We first apply a moving-average filter to smooth the posterior sequence, followed by a probability threshold to obtain frame-level decisions. To suppress spurious toggling caused by local acoustic fluctuations, a finite-state postprocessor enforces minimum voice and silence duration constraints, improving stability for both offline and streaming settings. Segments are optionally refined by merging short gaps, extending boundaries, and splitting overly long voice segments, which improves robustness for long-form audio and downstream ASR.

For mVAD, the same pipeline is applied independently to each event posterior stream (speech, singing, and music) with event-specific thresholds, yielding per-event timestamp segments. Non-streaming VAD outputs a set of voice segments with start/end timestamps; streaming VAD outputs incremental frame-level decisions and voice start/end events; mVAD outputs per-event timestamps for \textit{speech}, \textit{singing}, and \textit{music}, enabling event-aware downstream processing.

\section{FireRedLID: Hierarchical Spoken Language and Dialect Identification}
\label{sec:lid}

Spoken language identification (LID) \cite{valk2021voxlingua107,thukroo2022review,alashban2022spoken,o2025spoken} is a key component for multilingual and Chinese dialect speech processing in an all-in-one ASR system.
In practical deployments, LID is often used to route utterances to language-specific downstream processing, and errors in LID may propagate to subsequent modules such as ASR decoding and punctuation prediction.
FireRedLID is designed to be robust under diverse acoustic conditions and to support both multilingual language identification and fine-grained Chinese dialect identification.

\subsection{Model and training}
\label{sec:lid_method}

\noindent\textbf{Architecture}:
FireRedLID adopts an Encoder-Decoder-based architecture with a Conformer Encoder and a Transformer Decoder, following the implementation style of our AED ASR models.
Given an input utterance, the Encoder produces acoustic representations, and the Decoder generates a short token sequence that represents the LID result.

\noindent\textbf{Initialization}:
The Conformer Encoder is initialized from the pre-trained FireRedASR2-AED Encoder (\Cref{sec:asr2_aed}) to leverage its large-scale ASR representation learning.
The LID Decoder is randomly initialized and trained from scratch.

\noindent\textbf{Input features}:
FireRedLID uses the same acoustic features as FireRedASR2 (\Cref{sec:asr2}).

\noindent\textbf{Training data}:
FireRedLID is trained on approximately 200k hours of multilingual speech covering 100+ languages, including Mandarin and 20+ Chinese dialects.
The data is curated to include diverse domains and acoustic conditions to improve generalization.

\noindent\textbf{Training objective}:
FireRedLID is trained with a standard sequence-to-sequence cross-entropy objective using teacher forcing.

\subsection{Hierarchical label space and decoding}
\label{sec:lid_hier}

\noindent\textbf{Two-level labels}:
FireRedLID models LID as a two-level hierarchy.
The first level predicts the language (e.g., zh, en, ja, ko, etc.).
When the predicted language is Chinese (zh), the model additionally predicts a Chinese dialect label (e.g., mandarin, yue, wu, min, xiang, etc.).
This design reflects the natural label structure and improves stability for dialect identification by conditioning dialect prediction on the coarse language decision.

\noindent\textbf{Short-sequence token prediction}:
We formulate hierarchical LID as a short sequence generation task with a maximum decoding length of 2.
In practice, the model emits a language token first and typically emits a second token for Chinese dialect before generating <eos>.
For non-Chinese utterances, the decoder usually terminates after predicting the language token by emitting <eos>.
This formulation keeps the label sequence compact and reduces ambiguity compared with a flat label space.

\noindent\textbf{Decoding and confidence}:
During inference, we apply beam search to decode the label token sequence.
Since the output length is at most 2 tokens, decoding overhead is negligible.
We report the best hypothesis and compute an utterance-level confidence as the mean posterior probability of the decoded label tokens (excluding special tokens such as <sos> and <eos>).

\subsection{Supported languages and dialects}
\label{sec:lid_labels}

\noindent\textbf{Label coverage}:
FireRedLID supports 100+ languages and 20+ Chinese dialects.
We represent languages using compact language codes (e.g., zh, en, ja, ko) and group the 20+ Chinese dialects into 8 distinct geographical or linguistic dialect clusters (e.g., mandarin, yue, wu, min, etc.).
The complete lists of supported languages and dialects are provided in Appendix~\ref{app:lid_codes}.

\section{FireRedPunc: Punctuation Prediction}
\label{sec:punc}

FireRedPunc predicts punctuation for ASR transcripts to improve readability and downstream usability (e.g., subtitle display and machine translation).
It targets Chinese and English punctuation prediction for open-domain ASR outputs.

\noindent\textbf{Architecture}:
FireRedPunc adopts a BERT-style encoder \cite{devlin2019bert} with a token-level classification head.
Given an input token sequence, the model predicts a punctuation tag for each token, indicating the punctuation mark to be inserted after the token.
We initialize the encoder from a pre-trained LERT checkpoint \cite{cui2022lert}, a linguistically-motivated BERT variant, and fine-tune it for punctuation prediction.

\noindent\textbf{Punctuation set}:
We use a compact 5-way punctuation set corresponding to \texttt{no-punctuation} and the four marks \texttt{, . ? !}.
In our implementation, we use the Chinese full-width punctuation marks for Chinese text.
This design covers the most frequent punctuation marks in ASR applications while keeping the classifier simple and stable.

\noindent\textbf{Training data}:
FireRedPunc is trained on large-scale multi-domain text corpora with punctuation annotations.
The training data contains approximately 18.57B Chinese characters and 2.20B English words, covering diverse domains and writing styles to improve generalization to ASR-like inputs.

\noindent\textbf{Training objective}:
We train FireRedPunc with a standard token-level cross-entropy objective.

\noindent\textbf{Inference}:
At inference time, we tokenize ASR outputs using the same tokenizer as the pre-trained LERT encoder and apply the model to obtain token-level punctuation tags.
The final punctuated text is generated by inserting predicted punctuation marks into the original text sequence.

\section{Evaluation}
\label{sec:eval}

In this section, we evaluate FireRedASR2S on public benchmarks by reporting module-level results for ASR, VAD, LID, and punctuation prediction.
Each module is evaluated independently to avoid confounding effects introduced by upstream or downstream components.
Unless otherwise stated, all results are obtained in non-streaming settings.

\subsection{Evaluation of FireRedASR2}
\label{sec:eval_asr}

We evaluate FireRedASR2 on 24 public test sets covering Mandarin, Chinese dialects, and singing lyrics recognition.
FireRedASR2 includes two variants: FireRedASR2-LLM (8B+ parameters) and FireRedASR2-AED (1B+ parameters), representing different points on the accuracy-efficiency trade-off.

\noindent\textbf{Metric}:
We use Character Error Rate (CER, \%) for Chinese. Lower is better.
For aggregated results, we report:
(1) Avg-All-24: average CER across all 24 test sets,
(2) Avg-Mandarin-4: average CER across 4 Mandarin test sets,
(3) Avg-Dialect-19: average CER across 19 Chinese dialect test sets,
(4) Sing-1: CER on the singing lyrics test set (opencpop).
All averaged CERs are macro-averaged over test sets (equal weight per test set).

\noindent\textbf{Test sets}:
Avg-Mandarin-4 includes AISHELL-1 test set \cite{bu2017aishell} (aishell1), AISHELL-2 iOS test set\cite{du2018aishell} (aishell2), and WenetSpeech \cite{zhang2022wenetspeech} Internet/Meeting domains (ws-net/ws-meeting).
Avg-Dialect-19 includes KeSpeech \cite{tang2021kespeech} as well as dialect test sets curated from WenetSpeech-Yue \cite{li2025wenetspeech}, WenetSpeech-Chuan \cite{dai2025wenetspeech}, and MagicData \cite{magicdata} (see Appendix~\ref{app:asr_full_results} for the full list).
Sing-1 uses opencpop \cite{wang2022opencpop} for singing lyrics recognition.

\noindent\textbf{Baselines}:
We compare with strong commercial and open-source baselines:
(1) Doubao-ASR (commercial API) \cite{seedasr2024,vol2026doubaoasr},
(2) Qwen3-ASR (open-source checkpoint) \cite{shi2026qwen3},
(3) Fun-ASR (commercial API) \cite{an2025fun,ali2026funasr},
(4) Fun-ASR-Nano (open-source checkpoint; worse than Fun-ASR API; reported in Appendix) \cite{an2025fun}.
We emphasize that API-based baselines may change over time due to server-side updates and may include proprietary system-level components. We report their results as a practical reference point rather than a strictly reproducible baseline.
Full per-test-set results are provided in Appendix~\ref{app:asr_full_results}.

\begin{table}[h]
\caption{Comparison of Character Error Rate (CER\%) for FireRedASR2-LLM (FRASR2-LLM), FireRedASR2-AED (FRASR2-AED), and other large ASR baselines on public ASR test sets.}
\label{tab:eval_asr_main}
\centering
\setlength{\tabcolsep}{4pt}
\begin{threeparttable}
\begin{tabular}{l|ccccc}
\toprule
\textbf{Test set \textbackslash\ Model} & \textbf{FRASR2-LLM} & \textbf{FRASR2-AED} & \textbf{Doubao-ASR} & \textbf{Qwen3-ASR} & \textbf{Fun-ASR} \\
\midrule
Avg-All-24 & \textbf{9.67} & 9.80 & 12.98 & 10.12 & 10.92 \\
Avg-Mandarin-4 & \textbf{2.89} & 3.05 & 3.69  & 3.76  & 4.16 \\
Avg-Dialect-19  & \textbf{11.55} & 11.67 & 15.39 & 11.85 & 12.76 \\
Sing-1  & \textbf{1.12} & 1.17 & 4.36 & 2.57 & 3.05 \\
\midrule
aishell1    & 0.64 & 0.57 & 1.52 & 1.48 & 1.64 \\
aishell2    & 2.15 & 2.51 & 2.77 & 2.71 & 2.38 \\
ws-net      & 4.44 & 4.57 & 5.73 & 4.97 & 6.85 \\
ws-meeting  & 4.32 & 4.53 & 4.74 & 5.88 & 5.78 \\
\bottomrule
\end{tabular}

\begin{tablenotes}[flushleft]
\footnotesize
\item \textbf{API baselines:} Doubao-ASR (\texttt{volc.seedasr.auc}) was evaluated in early February 2026, and Fun-ASR was evaluated in late November 2025. API results may change over time due to server-side updates and may include proprietary components. To ensure a fair comparison, we disabled ITN and punctuation in the API outputs whenever such options were available, and used the default VAD configuration provided by each API.
\item \textbf{Data overlap:} Our ASR training data does not include any Chinese dialect or accented speech data from MagicData; all MagicData dialect datasets are used for evaluation only.
\end{tablenotes}
\end{threeparttable}
\end{table}

\noindent\textbf{Results and analysis}:
Table~\ref{tab:eval_asr_main} summarizes the main results.
FireRedASR2-LLM achieves the best overall accuracy across all aggregated metrics, reaching 2.89\% average CER on Mandarin (Avg-Mandarin-4), 11.55\% on Chinese dialect (Avg-Dialect-19), and 9.67\% on Avg-All-24. FireRedASR2 also performs strongly on singing lyrics recognition: on opencpop, FireRedASR2-LLM achieves 1.12\% CER.
FireRedASR2-AED achieves competitive accuracy with a smaller model size, providing a more balanced option for practical deployment.
Detailed per-test-set results are provided in Appendix~\ref{app:asr_full_results}.

\subsection{Evaluation of FireRedVAD}
\label{sec:eval_vad}

\noindent\textbf{Task and label definition}:
We evaluate FireRedVAD on a binary speech activity detection task (\textit{voice} vs.\ \textit{non-voice}).
This evaluation focuses on speech presence/absence and is aligned with typical VAD usage for ASR segmentation.

\noindent\textbf{Metrics}:
We report AUC-ROC, F1 score, False Alarm Rate (FAR), and Miss Rate (MR).
AUC-ROC is threshold-independent.
F1/FAR/MR depend on the decision threshold.

\noindent\textbf{Test set}:
We evaluate on FLEURS-VAD-102, a multilingual VAD benchmark covering 102 languages.
It is constructed by sampling approximately 100 audio files per language from the FLEURS test set and manually annotating binary VAD labels, resulting in 9,443 audio files in total. We will release FLEURS-VAD-102 and its annotation protocol to facilitate reproducible research.

\noindent\textbf{Frame setup}:
All VAD metrics are computed at the frame level.
We use 25ms analysis windows with a 10ms frame shift, consistent with the feature extraction setup used in FireRedASR2.

\noindent\textbf{Baselines and operating point}:
We compare with widely used open-source VAD systems, including Silero-VAD \cite{SileroVAD}, TEN-VAD \cite{TENVAD}, FunASR-VAD \cite{gao2023funasr}, and WebRTC-VAD \cite{webrtc}.
For threshold-dependent metrics (F1/FAR/MR), we use a fixed posterior threshold of 0.5 for all neural VAD models to provide a consistent operating point; tuning thresholds on a development set may further improve F1 for individual systems.

\begin{table}[!ht]
\caption{Frame-level VAD performance on FLEURS-VAD-102. Higher is better for AUC-ROC and F1; lower is better for FAR and MR.}
\label{tab:eval_vad}
\centering
\setlength{\tabcolsep}{4pt}
\begin{threeparttable}
\begin{tabular}{l|ccccc}
\toprule
\textbf{Metric \textbackslash\ Model} & \textbf{FireRedVAD} & \textbf{Silero-VAD} & \textbf{TEN-VAD} & \textbf{FunASR-VAD} & \textbf{WebRTC-VAD} \\
\midrule
AUC-ROC (\%) $\uparrow$ & \textbf{99.60} & 97.99 & 97.81 & -- & -- \\
F1 (\%) $\uparrow$      & \textbf{97.57} & 95.95 & 95.19 & 90.91 & 52.30 \\
FAR (\%) $\downarrow$   & \textbf{2.69}  & 9.41  & 15.47 & 44.03 & 2.83 \\
MR (\%) $\downarrow$    & 3.62           & 3.95  & 2.95 & 0.42 & 64.15 \\
\bottomrule
\end{tabular}
\begin{tablenotes}[flushleft]
\footnotesize
\item AUC-ROC is not reported for FunASR-VAD and WebRTC-VAD, as these systems do not output continuous posterior probabilities required for threshold-independent evaluation.
\end{tablenotes}
\end{threeparttable}
\end{table}

\noindent\textbf{Results and analysis}:
As shown in Table~\ref{tab:eval_vad}, FireRedVAD achieves strong multilingual VAD performance with 99.60\% AUC-ROC and 97.57\% F1, outperforming all compared baselines.
Notably, FireRedVAD achieves this SOTA performance with an exceptionally small parameter size ($\sim$0.6M), demonstrating a strong balance between accuracy and efficiency for practical industrial pipelines.
FireRedVAD maintains a low false alarm rate (2.69\%) while keeping a low miss rate (3.62\%), indicating a balanced operating point for downstream segmentation.
We note that some baselines (e.g., FunASR-VAD) achieve a very low miss rate but at the cost of a substantially higher false alarm rate, which may lead to excessive segmentation and unnecessary downstream ASR computation in practical deployments.

\subsection{Evaluation of FireRedLID}
\label{sec:eval_lid}

\noindent\textbf{Test sets}:
We evaluate FireRedLID on multilingual and Chinese dialect LID benchmarks.
For multilingual LID, we report results on the FLEURS \cite{fleurs2022arxiv} test set (82 languages) and CommonVoice \cite{ardila2020common} test set (74 languages).
For Chinese dialect identification, we evaluate on a combined benchmark by directly merging test samples from KeSpeech and MagicData, covering 10+ Chinese dialects.

\noindent\textbf{Metric}:
We report utterance-level LID accuracy (\%). Higher is better.

\noindent\textbf{Baselines}:
We compare with Whisper \cite{radford2023robust} language identification, SpeechBrain LID model \cite{speechbrain}, and Dolphin \cite{meng2025dolphin}.

\begin{table}[!ht]
\caption{Utterance-level LID accuracy (\%) on public test sets. Higher is better.}
\label{tab:eval_lid}
\centering
\begin{tabular}{l|c|c|c|c}
\toprule
\textbf{Test set \textbackslash\ Model} & \textbf{FireRedLID} & \textbf{Whisper} & \textbf{SpeechBrain} & \textbf{Dolphin} \\
\midrule
FLEURS test & \textbf{97.18} & 79.41 & 92.91 & --\\
CommonVoice test & \textbf{92.07} & 80.81 & 78.75 & -- \\
Chinese dialects & \textbf{88.47} & -- & -- & 69.01 \\
\bottomrule
\end{tabular}
\end{table}

\noindent\textbf{Results and analysis}:
Table~\ref{tab:eval_lid} shows that FireRedLID achieves strong performance on both multilingual and Chinese dialect LID tasks.
On FLEURS, FireRedLID reaches 97.18\% accuracy, substantially outperforming Whisper and improving over SpeechBrain.
On CommonVoice, FireRedLID also achieves the best accuracy among compared systems.
On the combined Chinese dialect benchmark, FireRedLID achieves 88.47\% accuracy, demonstrating the effectiveness of our hierarchical label modeling for fine-grained Chinese dialect identification.

\subsection{Evaluation of FireRedPunc}
\label{sec:eval_punc}

\noindent\textbf{Test sets}:
We evaluate FireRedPunc on internal multi-domain Chinese and English punctuation prediction benchmarks.
The Chinese benchmark contains 88,644 sentences, and the English benchmark contains 28,641 sentences. We will release the Chinese and English punctuation benchmarks to facilitate reproducible research.

\noindent\textbf{Label set and metric}:
FireRedPunc predicts punctuation tags from a compact set (space/no-punctuation, comma, period, question mark, and exclamation mark; see \Cref{sec:punc}).
For evaluation, we report Precision/Recall/F1 (\%) computed on punctuation labels.
According to our evaluation protocol, the Overall score is computed as micro-averaged Precision/Recall/F1 over all non-space punctuation labels.
Due to its extremely low frequency in the evaluation data, we exclude the exclamation mark from the reported Overall score.

\noindent\textbf{Baseline}:
We compare with a widely used punctuation model, FunASR-Punc (CT-Transformer) \cite{gao2023funasr}.

\begin{table}[!ht]
\caption{Punctuation prediction results on internal test sets (Precision/Recall/F1 in \%). Higher is better.}
\label{tab:eval_punc}
\centering
\begin{tabular}{l|c|c}
\toprule
\textbf{Test set \textbackslash\ Model}  & \textbf{FireRedPunc} & \textbf{FunASR-Punc} \\
\midrule
Multi-domain Chinese & \textbf{82.84 / 83.08 / 82.96} & 77.27 / 74.03 / 75.62 \\
Multi-domain English & \textbf{78.40 / 71.57 / 74.83} & 55.79 / 45.15 / 49.91 \\
Average F1 & \textbf{78.90} & 62.77 \\
\bottomrule
\end{tabular}
\end{table}

\noindent\textbf{Results and analysis}:
As shown in Table~\ref{tab:eval_punc}, FireRedPunc consistently outperforms the baseline on both Chinese and English benchmarks.
In particular, FireRedPunc achieves 82.96\% F1 on Chinese and 74.83\% F1 on English, resulting in a 78.90\% average F1.
The large gain on English suggests that our LERT-initialized BERT-style encoder and large-scale multi-domain training data are effective for punctuation prediction on ASR-like text.

\section{Discussion}
\label{sec:discussion}

We discuss key design choices and practical considerations of FireRedASR2S.

\noindent\textbf{System design: modularity with consistent interfaces}:
FireRedASR2S is designed as a modular pipeline consisting of VAD, LID, ASR, and punctuation prediction.
This design simplifies deployment and maintenance, allows independent iteration of each component, and improves reproducibility compared with ad-hoc integration of heterogeneous modules.

\noindent\textbf{Improved ASR accuracy and dialect coverage via data scaling}:
FireRedASR2 largely preserves the proven model architectures in FireRedASR and focuses on scaling supervised training data to approximately 200k hours with broader coverage.
The consistent improvements on Mandarin benchmarks and the strong performance on dialect test sets suggest that expanding supervised data diversity is a major driver for both recognition accuracy and generalization to diverse Chinese dialect scenarios.

\noindent\textbf{Human-labeled event supervision for segmentation}:
Compared to VAD models trained from ASR forced-alignment-derived supervision, FireRedVAD is trained on thousands of hours of human-annotated acoustic event data.
This explicit event supervision improves robustness under diverse acoustic conditions and supports both VAD and mVAD use cases.

\noindent\textbf{Hierarchical LID for languages and Chinese dialects}:
FireRedLID models LID as a short sequence generation task with hierarchical labels, predicting language first and dialect conditioned on Chinese.
This formulation better matches the label structure and reduces ambiguity compared with a flat label space, while keeping inference efficient.

\section{Conclusion}
\label{sec:conclusion}
We presented FireRedASR2S, a state-of-the-art industrial-grade all-in-one speech recognition system integrating ASR, VAD, LID, and punctuation prediction modules.
Building upon FireRedASR, FireRedASR2 improves recognition accuracy and expands coverage to a broader range of Chinese dialects, and provides two variants: an LLM-based model (8B+ parameters) for maximum accuracy and an AED-based model (1B+ parameters) for a balanced accuracy-efficiency trade-off.
FireRedVAD provides robust segmentation and achieves strong multilingual VAD performance.
FireRedLID performs hierarchical language and Chinese dialect identification with strong accuracy.
FireRedPunc restores punctuation for Chinese and English and achieves strong performance on multi-domain benchmarks.
We release model weights and code to facilitate research and practical deployment.
Future work will focus on further improving performance and expanding support for more languages.

\bibliographystyle{unsrt}
\bibliography{refs}

\clearpage
\appendix
\section*{Appendix}
\addcontentsline{toc}{section}{Appendix}

\section{Detailed ASR Results on Public Test Sets}
\label{app:asr_full_results}

\noindent
This appendix reports per-test-set CER(\%) on all 24 public test sets used in \Cref{sec:eval_asr}.
For completeness, we include Fun-ASR-Nano, which is the open-source checkpoint released by FunAudioLLM.

\begin{table}[h]
\caption{Comparison of Character Error Rate (CER\%) for FireRedASR2-LLM (FRASR2-LLM), FireRedASR2-AED (FRASR2-AED), and other large ASR baselines on public ASR test sets.}
\label{tab:eval_asr_full_24}
\centering
\setlength{\tabcolsep}{3pt}
\begin{threeparttable}
\small
\begin{tabular}{l|cc|cccc}
\toprule
\textbf{Test set \textbackslash\ Model} & \textbf{FRASR2-LLM} & \textbf{FRASR2-AED} & \textbf{Doubao-ASR} & \textbf{Qwen3-ASR} & \textbf{Fun-ASR} & \textbf{Fun-Nano} \\
\midrule
Avg-Mandarin-4 & \textbf{2.89} & 3.05 & 3.69 & 3.76 & 4.16 & 4.55 \\
Avg-Dialect-19 & \textbf{11.55} & 11.67 & 15.39 & 11.85 & 12.76 & 15.07 \\
Avg-All-24 & \textbf{9.67} & 9.80 & 12.98 & 10.12 & 10.92 & 12.81 \\
\midrule
aishell1          & 0.64 & 0.57 & 1.52 & 1.48 & 1.64 & 1.96 \\
aishell2          & 2.15 & 2.51 & 2.77 & 2.71 & 2.38 & 3.02 \\
ws-net            & 4.44 & 4.57 & 5.73 & 4.97 & 6.85 & 6.93 \\
ws-meeting        & 4.32 & 4.53 & 4.74 & 5.88 & 5.78 & 6.29 \\
\midrule
kespeech          & 3.08 & 3.60 & 5.38 & 5.10 & 5.36 & 7.66 \\
ws-yue-short      & 5.14 & 5.15 & 10.51& 5.82 & 7.34 & 8.82 \\
ws-yue-long       & 8.71 & 8.54 & 11.39& 8.85 & 10.14& 11.36\\
ws-chuan-easy     & 10.90& 10.60& 11.33& 11.99& 12.46& 14.05\\
ws-chuan-hard     & 20.71& 21.35& 20.77& 21.63& 22.49& 25.32\\
md-heavy          & 7.42 & 7.43 & 7.69 & 8.02 & 9.13 & 9.97 \\
md-yue-conv       & 12.23& 11.66& 26.25& 9.76 & 33.71& 15.68\\
md-yue-daily      & 3.61 & 3.35 & 12.82& 3.66 & 2.69 & 5.67 \\
md-yue-vehicle    & 4.50 & 4.83 & 8.66 & 4.28 & 6.00 & 7.04 \\
md-chuan-conv     & 13.18& 13.07& 11.77& 14.35& 14.01& 17.11\\
md-chuan-daily    & 4.90 & 5.17 & 3.90 & 4.93 & 3.98 & 5.95 \\
md-shanghai-conv  & 28.70& 27.02& 45.15& 29.77& 25.49& 37.08\\
md-shanghai-daily & 24.94& 24.18& 44.06& 23.93& 12.55& 28.77\\
md-wu             & 7.15 & 7.14 & 7.70 & 7.57 & 10.63& 10.56\\
md-zheng-conv & 10.20& 10.65& 9.83 & 9.55 & 10.85& 13.09\\
md-zheng-daily& 5.80 & 6.26 & 5.77 & 5.88 & 6.29 & 8.18 \\
md-wuhan          & 9.60 & 10.81& 9.94 & 10.22& 4.34 & 8.70 \\
md-tianjin        & 15.45& 15.30& 15.79& 16.16& 19.27& 22.03\\
md-changsha       & 23.18& 25.64& 23.76& 23.70& 25.66& 29.23\\
\midrule
opencpop          & 1.12 & 1.17 & 4.36 & 2.57 & 3.05 & 2.95 \\
\bottomrule
\end{tabular}

\begin{tablenotes}[flushleft]
\footnotesize
\item \textbf{Abbreviations:} ws denotes WenetSpeech; md denotes MagicData; conv denotes Conversational; daily denotes Daily-use; Fun-Nano denotes Fun-ASR-Nano-2512.
\item \textbf{API baselines:} Doubao-ASR (\texttt{volc.seedasr.auc}) was evaluated in early February 2026, and Fun-ASR was evaluated in late November 2025. API results may change over time due to server-side updates and may include proprietary components. To ensure a fair comparison, we disabled ITN and punctuation in the API outputs whenever such options were available, and used the default VAD configuration provided by each API.
\item \textbf{Data overlap:} Our ASR training data does not include any Chinese dialect or accented speech data from MagicData; all MagicData dialect datasets are used for evaluation only. The Fun-ASR API may benefit from proprietary training data, which could explain its advantage on certain dialect subsets (e.g., MagicData Shanghai and Wuhan dialect test sets).
\end{tablenotes}
\end{threeparttable}
\end{table}


\section{FireRedLID Label Lists}
\label{app:lid_codes}

\subsection{Full list of language codes}
\label{app:lid_lang_codes}
\begin{table}[!th]
\caption{Full list of language codes supported by FireRedLID.}
\label{tab:lid_lang_full}
\centering
\setlength{\tabcolsep}{3pt}
\renewcommand{\arraystretch}{1.05}
\begin{tabular}{c|l|l|c|l|l}
\toprule
Code & English Name & Chinese Name & Code & English Name & Chinese Name \\
\midrule
zh & Chinese & \chn{中文} & en & English & \chn{英语} \\
es & Spanish & \chn{西班牙语} & fr & French & \chn{法语} \\
ja & Japanese & \chn{日语} & ko & Korean & \chn{韩语} \\
ru & Russian & \chn{俄语} & de & German & \chn{德语} \\
pt & Portuguese & \chn{葡萄牙语} & ar & Arabic & \chn{阿拉伯语} \\
ab & Abkhazian & \chn{阿布哈兹语} & af & Afrikaans & \chn{南非荷兰语} \\
am & Amharic & \chn{阿姆哈拉语} & as & Assamese & \chn{阿萨姆语} \\
az & Azerbaijani & \chn{阿塞拜疆语} & ba & Bashkir & \chn{巴什基尔语} \\
be & Belarusian & \chn{白俄罗斯语} & bg & Bulgarian & \chn{保加利亚语} \\
bn & Bengali & \chn{孟加拉语} & br & Breton & \chn{布列塔尼语} \\
bs & Bosnian & \chn{波斯尼亚语} & ca & Catalan & \chn{加泰罗尼亚语} \\
ceb & Cebuano & \chn{宿务语} & cs & Czech & \chn{捷克语} \\
cy & Welsh & \chn{威尔士语} & da & Danish & \chn{丹麦语} \\
el & Greek & \chn{希腊语} & eo & Esperanto & \chn{世界语} \\
et & Estonian & \chn{爱沙尼亚语} & eu & Basque & \chn{巴斯克语} \\
fa & Persian & \chn{波斯语} & fi & Finnish & \chn{芬兰语} \\
fo & Faroese & \chn{法罗语} & gl & Galician & \chn{加利西亚语} \\
gn & Guarani & \chn{瓜拉尼语} & gu & Gujarati & \chn{古吉拉特语} \\
gv & Manx & \chn{马恩语} & ha & Hausa & \chn{豪萨语} \\
haw & Hawaiian & \chn{夏威夷语} & hi & Hindi & \chn{印地语} \\
hr & Croatian & \chn{克罗地亚语} & ht & Haitian Creole & \chn{海地克里奥尔语} \\
hu & Hungarian & \chn{匈牙利语} & hy & Armenian & \chn{亚美尼亚语} \\
ia & Interlingua & \chn{国际语} & id & Indonesian & \chn{印度尼西亚语} \\
is & Icelandic & \chn{冰岛语} & it & Italian & \chn{意大利语} \\
iw & Hebrew & \chn{希伯来语} & jw & Javanese & \chn{爪哇语} \\
ka & Georgian & \chn{格鲁吉亚语} & kk & Kazakh & \chn{哈萨克语} \\
km & Khmer & \chn{高棉语} & kn & Kannada & \chn{卡纳达语} \\
la & Latin & \chn{拉丁语} & lb & Luxembourgish & \chn{卢森堡语} \\
ln & Lingala & \chn{林加拉语} & lo & Lao & \chn{老挝语} \\
lt & Lithuanian & \chn{立陶宛语} & lv & Latvian & \chn{拉脱维亚语} \\
mg & Malagasy & \chn{马尔加什语} & mi & Māori & \chn{毛利语} \\
mk & Macedonian & \chn{马其顿语} & ml & Malayalam & \chn{马拉雅拉姆语} \\
mn & Mongolian & \chn{蒙古语} & mr & Marathi & \chn{马拉地语} \\
ms & Malay & \chn{马来语} & mt & Maltese & \chn{马耳他语} \\
my & Burmese & \chn{缅甸语} & ne & Nepali & \chn{尼泊尔语} \\
nl & Dutch & \chn{荷兰语} & no & Norwegian & \chn{挪威语} \\
oc & Occitan & \chn{奥克语} & pa & Punjabi & \chn{旁遮普语} \\
pl & Polish & \chn{波兰语} & ps & Pashto & \chn{普什图语} \\
ro & Romanian & \chn{罗马尼亚语} & sd & Sindhi & \chn{信德语} \\
si & Sinhala & \chn{僧伽罗语} & sk & Slovak & \chn{斯洛伐克语} \\
sl & Slovenian & \chn{斯洛文尼亚语} & so & Somali & \chn{索马里语} \\
sq & Albanian & \chn{阿尔巴尼亚语} & sr & Serbian & \chn{塞尔维亚语} \\
sv & Swedish & \chn{瑞典语} & sw & Swahili & \chn{斯瓦希里语} \\
ta & Tamil & \chn{泰米尔语} & te & Telugu & \chn{泰卢固语} \\
th & Thai & \chn{泰语} & tr & Turkish & \chn{土耳其语} \\
uk & Ukrainian & \chn{乌克兰语} & ur & Urdu & \chn{乌尔都语} \\
uz & Uzbek & \chn{乌兹别克语} & vi & Vietnamese & \chn{越南语} \\
yi & Yiddish & \chn{意第绪语} & yo & Yoruba & \chn{约鲁巴语} \\
\bottomrule
\end{tabular}
\end{table}

\subsection{Full list of Chinese dialect codes}
\label{app:lid_dialect_codes}
\begin{table}[!th]
\caption{Full list of Chinese dialect codes supported by FireRedLID.}
\label{tab:lid_dialect_full}
\centering
\begin{tabular}{c|p{6.5cm}|p{5.5cm}}
\toprule
Code & English Name & Chinese Name \\
\midrule
mandarin & Chinese (Mandarin) & \chn{中文 (普通话)} \\
yue & Chinese (Yue: Guangdong/Hong Kong) & \chn{中文 (粤语：广东/香港)} \\
wu & Chinese (Wu: Shanghai/Wu) & \chn{中文 (吴语：上海/吴语片区)} \\
min & Chinese (Min: Fujian) & \chn{中文 (闽语：福建)} \\
north &
Chinese (Mandarin-North: Shandong/Gansu/Ningxia/Hebei/Shanxi/Liaoning/Shaanxi) & \chn{中文 (官话-北方：山东/甘肃/宁夏/河北/山西/辽宁/陕西)} \\
xinan & Chinese (Mandarin-Southwest: Sichuan/Yunnan/Guizhou/Hubei/Chongqing) & \chn{中文 (官话-西南：四川/云南/贵州/湖北/重庆)} \\
xiang & Chinese (Xiang: Hunan) & \chn{中文 (湘语：湖南)} \\
bo & Tibetan (in Chinese context) & \chn{中文 (藏语)} \\
\bottomrule
\end{tabular}
\end{table}

\end{document}